\documentclass[12pt]{article}
\begin{document}

\begin{centering}
\vspace{.2in}
{\Large {\bf Critical points in two-dimensional replica sigma models}}\\
\vspace{.4in}
Paul Fendley${}$ \\
\vspace{.4in}
Department of Physics\\
University of Virginia\\
Charlottesville VA 22901, USA\\
\vspace{0.1in}
{fendley@virginia.edu}\\
\end{centering}
\vspace{.6in}
{\bf Abstract} \\ 

I survey the kinds of critical behavior believed to be exhibited in
two-dimensional disordered systems. I review the different replica
sigma models used to describe the low-energy physics, and discuss how
critical points appear because of WZW and theta terms.

\vspace{0.2in}

\vspace{.5in}


The last few years have seen a remarkable resurgence in activity on
disordered systems in two dimensions. Even though serious study in
this field goes back more than twenty years, recently there have been
a number of precise quantitative results.  The most famous
experimental problem of this sort is the transition between plateaus
in the quantum Hall effect. The experiments suggest strongly that
there is a critical point in between the plateaus, and all the
theoretical explanations of the quantum Hall effect indicate that
disorder plays a crucial role in this phase transition.

In this talk, I will discuss the kinds of critical behavior which
happen in systems of electrons which interact only with disorder.  It
is possible that electron interactions will substantially affect the
problem. However, until we theoretically understand the problem
without interactions, it is of course unlikely that we will be able to
understand the role of electron interactions.

In \cite{AZ, zirn} Hamiltonians bilinear in electron
annihilation/creation operators are classified.  The disorder appears
in the couplings of the Hamiltonian, i.e.\ the coefficients of the
various terms are taken to be random with some distribution.  For a
system with a finite number of states, each Hamiltonian is some matrix
belonging to the appropriate symmetry class (e.g. Hermitian, real
symmetric, etc.).  The simplest (zero-dimensional) way to analyze
these Hamiltonians would be to consider an ensemble of random matrices
belonging to the appropriate symmetry class. It turns out these
classes are related to symmetric spaces, so the random matrix
ensembles are conveniently labelled by a corresponding symmetric space
\cite{AZ,zirn}.  I present these labels in the first table, with a
brief physical description of the corresponding systems. For example,
models in the first three classes have a Hamiltonian describing
electrons with spin hopping on a lattice, with or without $SU(2)$
spin symmetry ${\cal S}$ and a discrete time-reversal symmetry ${\cal
T}$. Classes $A$III and $C$II describe electrons with no spin, but their
Hamiltonians allow Cooper pairing terms which simultaneously
annihilate or create a pair of electrons. The final four Hamiltonians
have spin and Cooper pairing terms.  The descriptions in parentheses
will be discussed later in this talk.
\bigskip\bigskip

\begin{center}
\begin{tabular}{|c||c|}\hline
RMT&description\\ \hline
$A$ (GUE)&Anderson localization, broken ${\cal S,T}$ (Hall plateau)\\
$A$I (GSE)&Anderson localization, broken ${\cal S}$, good ${\cal T}$\\
$A$II (GOE)&Anderson localization, good ${\cal S,T}$\\
\hline
$A$III&spinless superconductor, broken ${\cal T}$\\
$BD$I&\\
$C$II&spinless superconductor, good ${\cal T}$\\ 
\hline
$C$ & superconductor, broken ${\cal T}$, good ${\cal S}$ (SQHE, $d_{x^2-y^2}
+id_{xy}$)\\
$C$I& superconductor, good ${\cal T,S}$ ($d_{x^2-y^2}$)\\
$D$ &superconductor, broken ${\cal T,S}$ (random-bond Ising)\\
$D$III&superconductor, good ${\cal T}$, broken ${\cal S}$\\
\hline
\end{tabular}\\ \bigskip \end{center}
Table 1: Random matrix theories and some physical descriptions
\bigskip\bigskip

\noindent
It is worth noting that properties at non-zero frequency
$\omega\ne 0$ are described by the GOE, GSE or GUE universality
classes; the other universality classes only describe $\omega\to 0$
properties.

Since here I am interested in these systems in two spatial dimensions,
I will not utilize random matrix theory. However, because
classification according to random matrix ensembles is very convenient
for these sorts of Hamiltonians, most workers in this field still
refer to the systems by the names in the first column of table 1.
\bigskip

The disordered systems in table I can be described by certain sigma
models in the replica formulation. I review here what these words
mean.

Disorder is introduced into these fermion systems by allowing their
couplings to vary randomly with some distribution.  The action is
$S(\psi,R)$, where the fermions are denoted $\psi$ and the random
variables $R$.  The type of disorder considered here is called
quenched, meaning physical quantity is computed first for fixed
disorder, and then the randomness is averaged over. For example, the
free energy at fixed disorder $\ln Z(R)$ follows from the path
integral over the fermions for a given configuration $R$:
$$\ln Z(R)= \ln\left[ \int [{\cal D} \psi] e^{-S(\psi,R)}\right].$$
The physical free energy $\bar f$ is then given by averaging over the
random variables $R$:
\begin{equation}
\label{free}
\bar f = \int [{\cal D} R] \ln Z(R).
\end{equation}

The replica trick is a method for computing physical quantities which
seems to work at least some of the time. For the free energy, the
replica trick uses the identity
$$ \ln Z = \hbox{lim}_{N\to 0} \frac{Z^N-1}{N}$$ to rewrite $\ln Z(R)$
in (\ref{free}). The partition function to the $N^{\hbox{th}}$ power can
be written as the product of $N$ path integrals over $N$ ``replica''
fields $\psi_\mu$ where $\mu =1\dots N$, but all with the same random
fields $R$.  Then one makes the bold assumption that the $N\to 0$
limit and the $N+1$ path integrals commute, giving
$$\bar f = \hbox{lim}_{N\to 0} \frac{1}{N}\left[ -1 + \int [{\cal D}
R] \int [{\cal D} \psi_1] e^{-S(\psi_1,R)} \int [{\cal D} \psi_2]
e^{-S(\psi_2,R)} \dots \int[{\cal D} \psi_N]
e^{-S(\psi_N,R)}\right].$$ This trick thus puts the random fields on
the same footing as the replica fields, and one can use standard field
theory techniques on the problem.

The catch is that at the end of the computation one needs to set the
number of fields $N$ to be zero. Since rarely in a field theory does
one understand the analyticity properties of $N$, this step is
certainly not rigorous, or even close. Below I will take the strategy
of just doing computations at positive integer $N$ where everything is
well defined, and hope that the formulas are still valid when $N$ is
set to be zero. This procedure is well defined perturbatively, but we
will attempt to extract non-perturbative information as well. There is
another approach to these problems which utilizes supergroup
symmetries instead of the replica trick. This approach does not suffer
from all the possible ambiguities in taking this $N\to 0$
limit. However, it is not known how to utilize the techniques of
integrability in the supergroup formulation, so I am stuck with the
replica trick.

\bigskip

To make further progress on the problem we map the problem onto a
sigma model \cite{SW}.  The sigma model is an effective field theory
which follows (uniquely) by analyzing the symmetries of the replica
field theory. A sigma model is a field theory where the fields take
values on some manifold. In this talk, all of these manifolds are of
the form $G/H$, where $G$ and $H$ are Lie groups, with $H$ a maximal
subgroup of $G$. Such a manifold is called a symmetric space. The
best-known example of a sigma model on a symmetric space is often
called the sphere sigma model, where $G=O(3)$ and $H=O(2)$.  The field
can be thought of as a vector with three components and fixed length,
hence a sphere. While the vector can be rotated by the $O(3)$ group,
it is invariant under the $O(2)$ subgroup consisting of rotations
around its axis. Thus the space of distinct three-dimensional
fixed-length vectors (the two-sphere) is the coset $O(3)/O(2)$.

The replica sigma models describing the models in Table 1 can read off
the tables of \cite{zirn}. To see where they come from, it is useful
to do one example in detail. This discussion I steal from \cite{FK}.
The Hamiltonian describes spinless fermions with a
triplet p-wave type pairing:
\begin{equation}\label{eiii}
H_0 = \sum_k 
\epsilon_k c^\dagger_k c_k+(\Delta_k c^\dagger_k 
c^\dagger_{-k} + {\rm h.c.})
\end{equation}
where $\Delta_k \sim {v_\Delta \over 2}k_x$.
Disorder is weak enough to maintain some notion of the Fermi surface.
The low energy excitations of the fermions are then found about two
nodes positioned on the $k_y$-axis at $K_{\pm} = (0,\pm K)$.
Linearizing the theory about these
nodes via $\epsilon_{K_\pm + q} = q_y v_F$
and $c \sim c_1 \exp (iK_+ x) + c_2 \exp (iK_- x)$, $H$ becomes
\begin{equation}\label{eiv}
H = \int d^2x~ \psi^\dagger (iv_F \tau_z\partial_y
+ i v_\Delta \tau_x \partial_x)\psi .
\end{equation}
where $\psi^\dagger = (c^\dagger_{1},c_{2})$.  The Pauli matrices
$\tau_i$ act in the particle-hole space of the spinors.  $H$ is
precisely the Hamiltonian of a single Dirac fermion in $2+1$
dimensions. One can compute correlators
at fixed frequency $\omega$ by
utilizing the
action of a classical two-dimensional Euclidean field theory:
\begin{equation}\label{ev}
S_0 = \int d^2x~ \psi^\dagger(iv_F\tau_z\partial_y
+ i v_\Delta \tau_x\partial_x - i\omega \tau_z) \psi.
\end{equation}
One way of adding randomness is to include the term
\begin{equation}\label{evi}
H_{\rm disorder} = R (c^\dagger_1 c_1 + c^\dagger_2 c_2)
\end{equation}
in the Hamiltonian. Here $R$ is a random variable with variance
$\langle R(x) R(y)\rangle = u^{-1} \delta (x-y)$; because of the delta
function the disorder is called on-site. After replicating the
fermions, $\psi \rightarrow \psi_k$, the path integral over the random
field $R$ is easily done. This couples the different replicas via the
quartic terms $S_{\rm disorder} = -{1\over u}
(\psi^\dagger_k\tau^z\psi_k)(\psi^\dagger_l\tau^z\psi_l)$. Reorganizing
the fields via $\tilde\psi^\dagger \equiv
(\psi^\dagger_R,\psi^\dagger_L) =
\psi^\dagger\tau^z\exp(i\pi\tau^x/4)$ and $\tilde\psi \equiv (\psi_R,
\psi_L) = \exp(-i\pi\tau^x/4)\psi$ gives the action
\begin{equation}
S = \int d^2 x v_F\tilde\psi^\dagger_k(i\partial_y - 
\tau^z\partial_x)\tilde\psi_k
- {1\over u}(\tilde\psi^\dagger_k\tilde\psi_k)
(\tilde\psi^\dagger_l\tilde\psi_l).
\label{action}
\end{equation}
This theory is invariant under the group $U(N)_L\times
U(N)_R$ transforming $\tilde{\psi}\to \frac{1}{2}((1+\tau^z)U_L + (1-\tau^z)
U_R)\tilde{\psi}$.  Adding disorder to the Cooper-pairing term results in
another four-fermion term, but preserves this symmetry. As long as the
symmetry structure is unchanged, the low-energy physics should be
the same.

The sigma model describes the low-energy behavior of this system. The
idea is familiar from many different contexts in condensed-matter and
particle physics. The theory has a global symmetry $G$, which for this
example is $U(N)\times U(N)$. However, we assume there is an energy
scale where some fermion bilinear gets an expectation value. This
expectation value is invariant under only a subgroup $H$ of the full
symmetry group $G$. In dimensions higher than two, this would result
in spontaneous symmetry breaking. The sigma model describes the
interactions of the resulting Goldstone bosons, which take values on
the space $G/H$. In two dimensions, the symmetry does not break
spontaneously ($G$ remains a good global symmetry of the low-energy
theory), but still the low-energy degrees of freedom live on $G/H$.
The easiest way to see this explicitly is to introduce a
Hubbard-Stratonovich matrix field $M_{kl}$ to factor the
four-fermion term:
\begin{equation}\label{eviii}
S = \int d^2 x\, \left[iv_F\tilde\psi^\dagger_k(\partial_y - 
\tau^z\partial_x)\tilde\psi_k
- {1\over u}(\tilde\psi^\dagger_k M_{kl}\tilde\psi_l) +\hbox{tr}\, M^2\right]
\end{equation}
where $M$ is hermitian. Under the symmetry $U(N)_L\times U(N)_R$, $M
\to UMU^\dagger$.  The sigma model describes the physics around saddle points
of this path integral. For example, one can have
saddle points where $M$ is off-diagonal, e.g.\
$M_{LL} = M_{RR} = 0$, but $M_{LR} = M^\dagger_{RL}\propto I$, the
identity.  The diagonal subgroup $U(N)_V$ leaves this saddle point
invariant, and so the low-energy modes $T=M_{RL}$ take values in
$U(N)_L\times U(N)_R/U(N)_V \approx U(N)$. Focusing solely upon these
modes gives
\begin{equation}\label{eix}
S = S_0
- {1\over u}(\tilde\psi^\dagger_R T\tilde\psi_L
+ \tilde\psi^\dagger_L T^\dagger\tilde\psi_R).
\end{equation}
Integrating out the fermions leaves an effective action for the
bosonic field $T$, which can be expanded in in powers of the momentum
over the expectation value of $T$. I will discuss the form of
this action later. The important thing is that it follows solely from
the symmetry.

This model is in class $A$III in Table 1. This result can be read off
from the tables in \cite{zirn}. The replica sigma model corresponding
to a given disordered universality class is determined by taking $G/H$
to be the bosonic subspace labeled ``$M_F$'' in the second table in
\cite{zirn}. The $F$ is for fermion; we are using fermionic replicas
here. If we instead had used bosonic replicas to compute the same
Green's functions, we would have ended up with a replica sigma model
on a non-compact space; this model is labelled $M_B$ in Zirnbauer's
table. For the above example, this would be $Gl(N,C)/U(N)$. The
non-compact sigma model is not equivalent to the compact one except
hopefully in the replica limit $N\to 0$. However, very few exact
results are available for sigma models on non-compact spaces (in fact,
the difficulty in obtaining exact results for the supergroup sigma
models arises mainly from the non-compact bosonic subgroup, not from
the fermionic fields).

I will concentrate here exclusively on the replica sigma models on
compact spaces. I present a two-dimensional version of
Zirnbauer's tables in Table 2. The first column is the commonly-used
label coming from figure 1; I have rearranged the rows for reasons
which will become clear later. The second column is the replica sigma
model. The third column lists the types of critical behavior possible in
this sigma model. It is the purpose of the rest of this paper to
discuss the third column.

\begin{center}
\begin{tabular}{|c|c||c|}\hline
RMT&replica sigma model&possible 2D critical behavior\\ \hline
GUE&$U(2N)/U(N)\times U(N)$&Pruisken phase\\
$C$ &$Sp(2N)/U(N)$ &Pruisken phase\\
$D$ &$O(2N)/U(N)$ &Pruisken phase, metallic phase\\
\hline
$C$II&$U(N)/O(N)$ &$\theta=\pi \to U(N)_1$; Gade phase\\ 
GSE&$O(2N)/O(N)\times O(N)$&$\theta=\pi\to O(2N)_1$; metallic phase\\
\hline
$A$III&$U(N)\times U(N)/U(N)$ &WZW term; Gade phase\\
$C$I &$Sp(2N)\times Sp(2N)/Sp(2N)$ &WZW term\\
$D$III &$O(N)\times O(N)/O(N)$ & WZW term; metallic phase\\
\hline
$BD$I&$U(2N)/Sp(2N)$ &Gade phase?\\
GOE&$Sp(4N)/Sp(2N)\times Sp(2N)$&none!
\\
\hline
\end{tabular}\\ \smallskip \end{center}
Table 2: The replica sigma models and their possible critical behavior
\bigskip\bigskip

\noindent
The sphere sigma model we have discussed is the $N=1$ case of the
classes GUE, $C$, $C$II, and the $N$=2 case of classes $D$ and GSE (the
latter comprises two copies of the sphere).

It is important to note that even though each random matrix theory is
associated with a symmetric space as discussed in \cite{zirn} (this is
the origin of most of the names), this symmetric space is {\bf not}
necessarily the same as the symmetric space of the corresponding
replica sigma model.  For example, the random matrices for the theory
$C$I are in the tangent space of $Sp(2N)/U(N)$, (i.e.\ exponentiating
the random matrices gives the space $C$I$\equiv Sp(2N)/U(N)$), but the
corresponding replica sigma model is on the symmetric space
$Sp(2N)\times Sp(2N)/Sp(2N)$. 

Sigma models on symmetric spaces have the convenient property that
they have only one coupling constant. Roughly speaking, this means
that they preserve their ``shape'' under renormalization: only
their size changes. For example, the sphere sigma model remains a
sphere under renormalization: only the radius of the sphere
renormalizes. Actually, the sigma models for $A$III, $C$II and $BD$I
are not quite symmetric spaces: the corresponding symmetric spaces
have $SU(N)$ instead of $U(N)$. The extra $U(1)$ has some interesting
effects (for example turning a critical point into a critical line),
but does not affect the sigma model on the symmetric space at all.

At low energy, one can safely
neglect four-derivative terms and higher in the action. All
the above ``ordinary'' sigma models have an action
can be written in the form
\begin{equation}
S_{ordinary} = {g} \int dx\,dy\ 
\hbox{tr}\left[\partial_\mu T\,  \partial^\mu T^{-1}
\right]
\label{pcm}
\end{equation}
where $T$ is a unitary matrix, possibly with further restrictions.
This action is very non-trivial because of the non-linear constraint
of unitarity. For the example above, $T$ is an $N\times N$ unitary
matrix, because $U(N)\times U(N)/U(N) \approx U(N)$ as a space. For
$T$ in $U(N)/O(N)$ (class $C$II), $T$ is an $N\times N$ symmetric unitary
matrix. Symmetric matrices do not close under multiplication, which is
why $U(N)/O(N)$ is not a group but rather a coset. For
$O(2N)/O(N)\times O(N)$ (class GSE), $T$ is a $2N\times 2N$ symmetric
real orthogonal matrix.

An important observation of Anderson's is that the coupling constant
$g$ of the sigma model is related to the conductance of the system: if
$g$ renormalizes to zero, the system is localized and not
conducting, while if $g$ renormalizes to infinity, the system is
metallic. The latter is the trivial fixed point of the system. The
latter corresponds to the size of the manifold (e.g.\ the radius of
the sphere) going to infinity , which means the manifold becomes
effectively flat.  For ordinary sigma models on compact symmetric
spaces with $N>1$, the trivial fixed point is unstable.  The beta
function is proportional to the curvature of the manifold
\cite{friedan}, which is always positive for these spaces. Moreover
(for $N>1$), the beta function shows no evidence for a non-trivial
fixed point. This is a more or less a consequence of the
Mermin-Wagner-Coleman theorem, which says that continuous symmetries
can not be spontaneously broken in two dimensions. This implies that
the manifold should be always curved (and not critical); otherwise
there would be Goldstone modes in violation of the theorem.

However, in two dimensions, there are a number of
different ways critical behavior appears in a disordered system. I
group them into three types:
\begin{list}
\medskip
\item{1.} Perturbative Peculiarities
\subitem{a)} Gade phase
\subitem{b)} Metallic phase
\medskip
\item{2.} WZW term
\medskip
\item{3.} Theta term
\subitem{a)} Pruisken type (non-vanishing $\sigma_{xy}$)
\subitem{b)} $\theta = 0$ or $\pi$ only
\end{list}

\medskip
\noindent
We will discuss all these methods, but devote most of our attention to
2 and 3. These require adding extra terms to the action $S_{ordinary}$.

\bigskip\medskip

\noindent
1. {\bf Perturbative Peculiarities}
\medskip

While it is important to note that the Mermin-Wagner-Coleman theorem
implies no non-trivial fixed points in ordinary sigma models for
$N>1$, this of course does not prohibit interesting things from
happening in the replica limit $N\to 0$ \cite{mckane}. For example,
the dimension of all the manifolds in table II goes to zero as $N\to
0$, making the idea of positive curvature somewhat confusing. In fact,
a while ago (see \cite{hikami} and references within) it was shown
that at one loop
\begin{eqnarray}
\nonumber
\beta &\propto  N \qquad\qquad &\hbox{ if } G=U(N)\\
\nonumber
\beta &\propto  N-2 \qquad& \hbox{ if } G=O(N)\\
\nonumber
\beta &\propto  N+1 \qquad& \hbox{ if } G=Sp(2N)
\end{eqnarray}
where the constant of proportionality is a negative number for $N\ge
0$.  Therefore, by ``perturbative peculiarities'' I mean the
consequences of the fact that as $N\to 0$ the $\beta$ function goes
to zero for $G=U(N)$ and has changed sign for $G=O(N)$.

If the beta function goes to zero to all orders as $N\to 0$, this
opens up the possibility that there is no flow in the sigma model: for
any $g$ the model is at a fixed point. This indeed happens for classes
$A$III and $C$II (and probably $BD$I, although I am not aware of any
explicit computations other than of the perturbative beta
function). This behavior was discussed at length in \cite{Gade}, which
is why I call this the Gade phase. In Gade's work these universality
classes are realized by a particle hopping on a bipartite lattice (the
particle is restricted to hop only from one sublattice to the other);
class $C$II has time-reversal symmetry, while class $A$III does
not. The supergroup approach to these models was discussed in detail
in \cite{GLL}. Since these models also have an extra $U(1)$ factor as
mentioned above, the model is critical over an entire plane of
couplings.

If the beta function is positive as $N\to 0$, the trivial fixed point
is stable. This happens for classes $D$, $D$III and GSE. Since $g$
renormalizes to infinity, this phase is conducting and is hence called
metallic.  This implies the existence of at least one non-trivial
fixed point, because it is still expected that for strong enough
disorder, the system does not conduct. Hence there should be a phase
transition from the metallic phase to a localized phase at some value
of $g$. This fixed point should be unstable in $g$ in both directions
(i.e.\ for $g<g_c$ the system renormalizes to $g\to 0$, while for
$g>g_c$ the system renormalizes to $g\to\infty$).

\bigskip\medskip
\noindent
2. {\bf WZW term}
\medskip

Two-dimensional sigma models with a manifold of the form
$H\times H/H$ are called
principal chiral models, and have been widely studied. They are
massive asymptotically-free field theories for $N>1$. However, there
is an additional term which can be added to the action (\ref{pcm})
which changes the low-energy behavior from gapped to gapless for any
$N$. This is called the Wess-Zumino-Witten term. To
write it out explicitly, first one needs to consider field configurations
$h(x,y)$ which fall off at spatial infinity, so that one can take
the spatial coordinates $x$ and $y$ to be on a sphere.
Then one needs to extend the fields $h(x,y)$ on
the sphere to fields $h(x,y,z)$ on a ball which has the original
sphere as a boundary. The fields inside the ball are defined so that
$h$ at the origin is the identity matrix, while $h$ on the boundary is the
original $h(x,y)$. It is possible to find a continuous deformation of
$h(x,y)$ to the identity because $\pi_2(H)=0$ for any simple Lie group.
Then the WZW term is $k\Gamma$, where
\begin{equation}
\Gamma= \frac{\epsilon_{abc}}{24\pi}
\int dx dy dz\, \hbox{tr}\left[(h^{-1}\partial_a h) 
(h^{-1}\partial_b h) (h^{-1}\partial_c h) \right].
\label{wzw}
\end{equation}
The coefficient $k$ is known as the level, and for compact groups must
be an integer because the different possible extensions of $h(x,y)$ to
the ball yield a possible ambiguity of $2n\pi$ in $\Gamma$.  The WZW
term changes the equations of motion and beta function, but only by
terms involving $h(x,y)$: the variation of the integrand is a total
derivative in $z$.

The two-dimensional sigma model with WZW term has a stable fixed point
at $1/g=16\pi/k$ \cite{WZW}, so the model is critical and the
quasiparticles are gapless.  The corresponding con\-for\-mal field
theory is known as the $H_k$ WZW model \cite{KZ}.  The WZW term is
invariant under discrete parity transformations (e.g.\ $x\to -x$,
$y\to y$) if $h\to h^{-1}$ under this transformation.  For there to be
a WZW term in a parity-invariant theory, some of the low-energy fields
must be pseudoscalars.

The WZW term allows some of these sigma models which has a
stable low-energy fixed point. Why this term must arise in many
situations was understood in particle physics some time ago. This was
the topic of \cite{FK}.
The WZW term was shown to arise in several disordered systems
\cite{NTW,BSZ}, by bosonizing the explicit system. There it appears to
ensure the certain current-algebra commutation relations are obeyed
properly in the bosonic theory. It appears more generally in the map
of the action like (\ref{eviii}) to a sigma model. The sigma model
arises when integrating out the fermions in this action, leaving a
low-energy effective action for $h$. Upon doing so, one easily obtains
the ordinary sigma model action of the form (\ref{pcm}) (with
$T=h$). The WZW term arises for a subtler reason. To perform a
consistent low-energy expansion of the action , one must change field
variables. This results in a Jacobian in the path integral \cite{Fuj},
which in these two-dimensional cases is precisely the WZW term
(\ref{wzw}).

In fact, one can determine without these explicit computations whether
or not the WZW term will appear in the low-energy effective action.
The reason is some very deep physics known as the chiral anomaly.  In
the models which admit a WZW term, the fermions have a chiral
$H_L\times H_R$ symmetry.  As is well known, chiral symmetries
involving fermions are frequently anomalous.  Noether's theorem says
that a symmetry of the action gives a conserved current with
$\partial_\mu j^\mu=0$, but this is only true to lowest order in
perturbation theory. An anomaly is when the current is not conserved
in the full theory (although the associated charge is still
conserved).  In the case of massless fermions in $1+1$ dimensions,
this was shown in detail in \cite{CGJ,WZW}.  The WZW term is the
effect of the anomaly on the low-energy theory. Even though the
fermions effectively become massive when $T$ gets an expectation
value, their presence still has an effect on the low-energy theory,
even if this mass is arbitrarily large.  This violation of decoupling
happens because the chiral anomaly must be present in the low-energy
theory. In other words, the anomaly coefficient does not
renormalize. This follows from an argument known as 't Hooft anomaly
matching \cite{Hooft}. One imagines weakly gauging the anomalous
symmetry. It is not possible to gauge an anomalous symmetry in a
renormalizable theory, but one can add otherwise non-interacting
massless chiral fermions to cancel the anomaly. Adding these spectator
fermions ensures that the appropriate Ward identities are obeyed, and
the symmetry can be gauged. In the low-energy effective theory, the
Ward identities must still be obeyed. Because the massless spectator
fermions are still present in the low-energy theory, there must be a
term in the low-energy action which cancels the anomaly from the
spectators. This is the WZW term.

To determine whether a chiral anomaly and hence a WZW term is present,
one usually needs to do only a simple one-loop computation. For chiral
symmetries, it is customary to define the vector and axial currents
$j^\mu_V = j^\mu_L+j^\mu_R$, $j^\mu_A = j^\mu_L-j^\mu_R$; for these
theories $j^\mu_A=\epsilon^{\mu\nu}j_{V\nu}$. One then computes the
correlation functions $\langle j_{V}^\mu(x,y)
j_{V}^\nu(0,0)\rangle\equiv C^{\mu\nu}$. If $\partial_\mu
C^{\mu\nu}\ne 0$ or $\epsilon_{\alpha\beta}\partial^\alpha
C^{\beta\nu}\ne 0$ for some $\nu$, then there is an anomaly. An
important characteristic of the anomaly is that it is independent of
any continuous change in the theory, as long as the chiral symmetry is
not broken explicitly. Thus to find the anomaly, one can compute
$C^{\mu\nu}$ using free fermions (where the only contribution is a
simple one-loop graph, see e.g. \cite{WZW}).

For the $p$-wave superconductor example discussed in detail above
(class $A$III), one has $H=U(N)$ and the possibility of a WZW
term. With two nodes in the fermi surface as discussed above, one in
fact does, with $k=1$ \cite{DF,FK}. The model is already
critical because it is in a Gade phase, but 
the WZW term does have an effect; for example when the
coupling $g$ is at the WZW fixed point, the density of states falls
off with a power law in energy instead of an exponential \cite{GLL}.

Classes $C$I and $D$III are more interesting in this context.  

Class $C$I basically amounts to generalizing (\ref{eiii}) to include the
spin of the electron, and preserving the $SU(2)$ spin symmetry
\cite{Senthil}. As discussed in \cite{NTW,FK}, if there are two nodes
in the fermi surface, one ends up with an $Sp(2N)_1=U(N)_2$ critical
point. The case of most interest is the $d_{x^2-y^2}$ superconductor, where
there are four nodes in the Fermi surface. If the two sets of nodes
are not coupled then the global symmetry is enlarged to $Sp(2N)\times
Sp(2N)\times Sp(2N)\times Sp(2N)$. 't Hooft's argument means that
one obtains two copies of the $Sp(2N)_1$ theory with conserved
currents
$$\partial_\mu j^\mu_{V1} =\partial_\mu j^\mu_{V2} =0$$
$$\partial_\mu j^\mu_{A1}= \partial_\mu j^\mu_{A2} \ne 0.$$ The reason
the anomalies in the second line are the same is that I have defined
the two sets of fields with the same conventions (i.e.\ the same sign
WZW term). However, generic disorder couples all the nodes, breaking
the symmetry back to $Sp(2N)\times Sp(2N)$. If the remaining
symmetries are generated by $j^\mu_{V1}+j^\mu_{V2}$ and
$j^\mu_{A1}+j^\mu_{A2}$, then the latter is anomalous. Because of 't
Hooft's argument, the two anomalies just add and one obtains
$Sp(2N)_2$ \cite{FK}. However, the conserved symmetries of the
$d_{x^2-y^2}$ superconductor are determined by the band structure, and
this requires that the conserved currents with these conventions be
$j^\mu_{V1}+j^\mu_{V2}$ and $j^\mu_{A1}-j^\mu_{A2}$ \cite{ASZ}. The
anomalies in the latter cancel, and therefore there is no WZW term and
no phase transition in the resulting sigma model. I have used
relativistic notation in this section, but even if 1+1 dimensional
Lorentz invariance is broken (e.g. by taking $v_{F1}/v_{\Delta1} \ne
v_{F2}/v_{\Delta2}$), these results still apply.

A WZW term likewise appears in class $D$III when there are two
nodes. For $N>2$ this fixed point is stable like the others. As
discussed above, the sign of the beta function has flipped as $N\to
0$, so the model has a metallic phase. This makes the $N\to 0$ limit
of the $O(2N)_k$ fixed point an excellent candidate for the unstable
fixed point between the metallic and localized phases. The subtlety
here is that it is not clear whether or not one can continue the
well-understood $N>2$ results to the replica
limit. There is at least one well-known situation (two-dimensional
self-avoiding polymers) where one cannot do such a continuation. I am
currently studying this question, but do not yet have any conclusive
answers.

There are two morals of this section:

\begin{list}
\medskip
\item{1.} When the WZW term is present, the model 
has a non-trivial fixed point.
\item{2.} Whether or not the WZW term is present depends on original
(microscopic) disordered model considered. For the case of fermions,
the coefficient of this term is easily determined.

\end{list}
\bigskip\medskip
\noindent
3. {\bf $\theta$ term}
\medskip

Another term which can be added to some sigma model actions is called
the theta term. This term is inherently non-perturbative: it does not
change the beta function derived near the trivial fixed point at
all. Nevertheless, it can result in a non-trivial fixed point.

The theta term is best illustrated in the sphere sigma model.  As
above, I consider field configurations which go to a constant at
spatial infinity, so the spatial coordinates are effectively that of a
sphere.  Since the field takes values on a sphere, the field is
therefore a map from the sphere to a sphere. An important
characteristic of such maps is that they can have non-trivial
topology: they cannot necessarily be continuously deformed to the
identity map. This is analogous to what happens when a circle is
mapped to a circle (i.e.\ a rubber band wrapped around a pole): you
can do this an integer number of times called the winding number (a
negative winding number corresponds to flipping the rubber band upside
down). It is the same thing for a sphere: a sphere can be wrapped
around a sphere an integer number of times. An example of winding
number 1 is the isomorphism from a point on the spatial sphere to
the same point on the field sphere. The identity map is
winding number 0: it is the map from every point on the spatial
sphere to a single point on the field sphere, i.e. $T(x,y)=$constant.
In field theory, field configurations with non-zero winding number are
usually called instantons. The name comes from viewing one of the
directions as time (in our case, one would think of say $x$ as space
and $y$ as Euclidean time). Since instanton configurations fall off to
a constant at $y=\pm\infty$, the instanton describes a process local
in time and hence ``instant''. In an even fancier modern language, an
instanton would be called a $-1$ brane. Therefore, the field
configurations in the sphere sigma model can be classified by an
integer $n$. We can therefore add a term $$S_\theta = in\theta$$ to
the action, where $\theta$ is an arbitrary parameter. Since $n$ is an
integer, the physics is periodic under shifts of $2\pi$ in $\theta$.

Pruisken \cite{Pruisken} showed that the replica sigma model
describing two-dimensional non-interacting electrons with disorder and
a strong transverse magnetic field (which breaks ${\cal T}$) is in the
GUE class. This model has instantons with integer winding number, and
so allows a $\theta$ term. He showed that while the sigma model
coupling $g$ is related to the conductivity $\sigma_{xx}$, the other
parameter $\theta$ is related to the Hall conductivity
$\sigma_{xy}$. He proposed that at $\theta=\pi$, the system has a
critical point separating a phase with $\sigma_{xy}=0$ from
$\sigma_{xy}=1$: the famous (experimentally observed) transition
between quantum Hall plateaus. This critical point is stable in $g$
but unstable when $\theta$ is taken away from $\pi$. While Pruisken's
proposed phase diagram is widely believed to be correct, noone has
succeeded in deriving any analytic results valid for the replica limit
$N\to 0$.  The best evidence that Pruisken's phase diagram does apply
to GUE class models is indirect, coming from numerical studies of the
network model \cite{Chalker}. The network model can be mapped on to a
supergroup spin chain \cite{nick} whose continuum limit should be
described by a supergroup sigma model of the GUE class. Thus ven
though the network model is microscopically different from the model
of electrons with disorder and transverse magnetic field, it should be
in the same universality class. Numerical studies are much easier to
do on the network model or on the supergroup spin chain, and the work
done is completely consistent with this phase diagram.

Although we do not have any exact results applicable the GUE replica
limit, we do a number of exact results for some sigma models at
$\theta=\pi$, and I would like to discuss them in this section.  These
all are in harmony with Pruisken's picture.  Although Pruisken did not
know this at the time of his proposal (he was reasoning by analogy
with some of 't Hooft's work on QCD \cite{oblique}), the sphere sigma
model has the same structure as he proposed for the Hall
plateaus. Another way of saying this would be to say the phase
structure of the $U(2N)/U(N)\times U(N)$ sigma model is believed to be
the same for $N=1$ and $N\to 0$. While we do not know the exact nature
of the non-trivial critical point for $N\to 0$, it is understood well
for the sphere, $N=1$. Namely, Haldane realized when studying the
half-integer-spin Heisenberg spin chains that the sphere sigma model
at $\theta=\pi$ has a non-trivial fixed point stable in $g$. This
fixed point turns out to be exactly the $SU(2)_1$ WZW model
\cite{Affleck}. The argument goes as follows. First one uses
Zamolodchikov's c-theorem, which makes precise the notion that as one
follows renormalization group flows, the number of degrees of freedom
goes down. Zamolodchikov shows that there is a quantity $c$ associated
with any two-dimensional unitary field theory such that $c$ must not
increase along a flow. At a critical point, $c$ is the central charge
of the corresponding conformal field theory \cite{cthm}. At the
trivial fixed point of a sigma model where the manifold is flat, the
central charge is the number of coordinates of the manifold (i.e.\ the
number of free bosons which appear in the action (\ref{pcm})). For the
sphere, this means that $c=2$ at the trivial fixed point. The only
unitary conformal field theories with $SU(2)$ symmetry and $c<2$ are
$SU(2)_k$ for $k<4$ (in general, the central charge of $SU(2)_k$ is
$3k/(k+2)$). One can use the techniques of \cite{KZ} to show that
there are relevant operators at these fixed points, and at $k=2$ or
$3$, no symmetry of the sphere sigma model prevents these relevant
operators from being added to the action \cite{Affleck}. So while it
is conceivable that the sphere sigma model with $\theta=\pi$ could
flow near to these fixed points, these relevant operators would
presumably appear in the action and cause a flow away. However, there
is only one relevant operator (or more precisely, a multiplet
corresponding to the field $h$ itself) for the $SU(2)_1$ theory. The
sigma model has a discrete symmetry $T\to -T$ when $\theta=\pi$; the
winding number $n$ goes to $-n$ under this symmetry, but $\theta=\pi$
and $\theta=-\pi$ are equivalent because of the periodicity in
$\theta$. This discrete symmetry of the sigma model presumably turns
into the symmetry $h\to -h$ of the WZW model. While the operator
$\hbox{tr}\, h$ is $SU(2)$ invariant, it is not invariant under this
discrete symmetry. Therefore, the only possible low-energy fixed point
for the sphere sigma model at $\theta=\pi$ is $SU(2)_1$. A variety of
arguments involving the spin chain suggest strongly that this does in
fact happen \cite{Affleck}.

This picture also shows what happens when $\theta$ is moved away from
$\pi$. Here the discrete symmetry is broken but $SU(2)$ is preserved,
so one adds $\hbox{tr}\, h$ to the $SU(2)_1$ action. This is relevant,
and in fact is equivalent to the sine-Gordon model (with dimension 1/2
$\cos$ term). This is a massive field theory, with no non-trivial
low-energy fixed point. The sphere sigma model reproduces exactly
Pruisken's phase diagram.

The flow of the sphere sigma model at $\theta=\pi$ to the $SU(2)_1$
WZW model was essentially proven in \cite{ZZ}.  This result does not
seem to be widely known, so I will review it here.  The sphere sigma
model is integrable at $\theta=0$, meaning that there are an infinite
number of conserved currents which allow one to find exactly the
spectrum of quasiparticles and their scattering matrix in the
correspoding $1+1$ dimensional field theory. There is evidence that
$\theta=\pi$ case is integrable also, so one can assume so and go on
to find the quasiparticle $S$ matrix here as well. This is done in
\cite{ZZ}. They find that while the quasiparticles for $\theta=0$ are
gapped and form a triplet under the $SU(2)$ symmetry, for $\theta=\pi$
they are gapless, and form $SU(2)$ doublets (left- and
right-moving). This is a beautiful example of charge
fractionalization: the field $T$ is a triplet under the $SU(2)$
symmetry, but when $\theta=\pi$ the excitations of the system are
doublets. They compute the $c$ function, and find that at high energy
$c$ indeed is $2$ as it should be at the trivial fixed point, while
$c=1$ as it should be at the $SU(2)_1$ low-energy fixed point. As an
even more detailed check, the free energy at zero temperature in the
presence of a magnetic field was computed for both $\theta=0$ and
$\pi$ \cite{sausage}. The results can be expanded in a series around
the trivial fixed point. One can identify the ordinary pertubative
contributions to this series, and finds that they are the same for
$\theta=0$ and $\pi$, even though the particles and $S$ matrices are
completely different. This is as it must be: instantons and hence the
$\theta$ term are a boundary effect and hence cannot be seen in
ordinary perturbation theory. One can also identify the
non-perturbative contributions to these series, and see that they
differ. Far away from the trivial fixed point, non-perturbative
contributions can dominate, which is why $\theta=0$ has no non-trivial
fixed point, while $\theta=\pi$ does.

The question now is if similar behavior is found for any other
disordered universality classes in two dimensions. In the sigma model
language, the question is if any other of the models in Table 2 have
instantons and hence allow a $\theta$ term. This question has already
been answered by mathematicians; for a review accessible to
physicists, see \cite{sidney}. In mathematical language, the question
is whether the second homotopy group $\pi_2(G/H)$ is non-trivial. The
second homotopy group is just the group of winding numbers of maps
from the sphere to $G/H$, so for the sphere it is the integers. The
general answer is that $\pi_2(G/H)$ is the kernel of the embedding of
$\pi_1(H)$ into $\pi_1(G)$, where $\pi_1(H)$ is the group of winding
numbers for maps of the circle into $H$. We have seen already that
$\pi_1(H)$ is the integers when $H$ is the circle $=U(1)=SO(2)$. The
only simple Lie group $H$ for which $\pi_1$ is nonzero is $SO(N)$, where
$\pi_1(SO(N))={\bf Z}_2$ for $N\ge 3$ and ${\bf Z}$ for $N$=2. Thus
there are models with integer winding number, some with just winding
number $0$ or $1$, and some with no instantons at all. Integer winding
number means that $\theta$ is continuous and periodic, while a winding
number of $0$ or $1$ means that $\theta$ is just $0$ or $\pi$ (just
think of $\theta$ as being the Fourier partner of $n$).  The results
are summarized in the last column of Table 2; the models with integer
winding number are labelled as having a Pruisken phase, while those
with ${\bf Z}_2$ winding number are $C$II and GSE.

The replica sigma models with integer winding number and continuous
$\theta$ are believed to behave like the sphere sigma model.  In
addition to the GUE class, this happens for the $Sp(2N)/U(N)$ sigma
models (class $C$) and the $O(2N)/U(N)$ sigma models (class $D$). The
replica limit of class $C$ should have Pruisken's phase diagram, while
in class $D$ it should be modified because of the flip in sign of the
beta function: the non-trivial fixed point at $\theta=\pi$ should be
unstable in $\theta$ and $g$, and another non-trivial fixed point
should appear at some value $g_c,\theta=0$ (because the metallic phase
should not exist at small enough $g$, i.e. strong enough disorder).

In all three of these universality classes, there is a network model
(roughly speaking, one for each type of simple Lie group, $U(N)$,
$Sp(2N)$ and $O(2N)$) \cite{Chalker,SQHE,CR}. In all three cases,
numerical results on the network model are consistent with the
existence of a non-trivial fixed point as Pruisken predicted.  Class
$D$ turns out to be a complicated story \cite{Senthil,RG,BSZ,CR,GLR}. It
seems that the sigma model does not describe all the physics of this
class: because of the existence of domain walls, the complete phase
diagram involves more than the two parameters $\theta$ and $g$ of the
sigma model and Pruisken's phase diagram \cite{BSZ,GLR}. The
two-dimensional random-bond Ising model belongs to this symmetry
class, but is a subspace of this full space.  All results
support the existence of a non-trivial critical point (or actually,
points), but very little is known about detailed properties.  On the
other hand, the class $C$ model, known as the spin quantum Hall effect
(SQHE) \cite{SQHE}, is better understood. There is a remarkable exact
result mapping certain correlators at the non-trivial fixed point onto
a known conformal field theory, that describing percolation
\cite{ilya}. There are no exact results from the replica sigma model
point of view, but I have a conjecture to which I will return below.

It is not yet known whether the sigma models in the GUE, $C$ or $D$
classes are integrable. I do have a variety of exact results for the
models with ${\bf Z}_2$ instantons, namely classes $C$II and GSE
\cite{me}. Without a continuous $\theta$ parameter, there does not
seem to be any $\sigma_{xy}$, so these models do not have the full
structure of the above three classes. However, these two models still
have a non-trivial fixed point when $\theta=\pi$, and for this reason
I believe they provide strong support for Pruisken's picture.

Basically, my results generalize those of \cite{ZZ} to
this much more general case. This is important because to have any
hope of being able to take the replica limit, one needs a solution for
any $N$.  For class $C$II, the sigma models are on the space
$U(N)/O(N)$. This sigma model has action (\ref{pcm}) with $T$ a
symmetric, unitary matrix.  I find that when $\theta=0$, these sigma
models are the $U(N)$ generalization of the sphere sigma model.  When
$\theta=0$, the model has a gap, with the spectrum consisting of
massive particles in the symmetric representation of $SU(N)$ (plus
bound states in more general representations). When $\theta=\pi$, the
spectrum consists of gapless quasiparticles which are in the
fundamental representations (vector, antisymmetric tensor,\dots) of
$SU(N)$. The non-trival low-energy fixed point when $\theta=\pi$
corresponds to $SU(N)_1\times U(1)$. Thus we see that at least for
$N>0$, the replica sigma models in class $C$II with $\theta=\pi$ have
exactly the same fixed point as those in class $A$III when a $k=1$ WZW
term is present! The effect of having $\theta=\pi$ is as
discussed before: the density of states changes its behavior near
special values of the coupling.

The results for the GSE class are similar. This model is the
$O(2N)/O(N)\times O(N)$ sigma model, which has action (\ref{pcm}) with
$T$ a symmetric, real and orthogonal matrix. When $\theta=0$, the
model is gapped. When $\theta=\pi$, the model is gapless with a
non-trivial stable fixed point corresponding to the $O(2N)_1$ WZW
model. This sigma model proves to be the $O(2N)$ generalization
of the sphere sigma model. The $O(2N)_1$ model turns out to be $2N$
free Majorana fermions. The word ``free'' is slightly deceptive,
because just as in the 2d Ising model, one can study correlators of
the magnetization or ``twist'' operator, which are highly non-trivial.
Because of the changes in the beta function at $N$=1, it is not clear
yet whether these results can be continued to the replica limit; I am
currently studying this. However, again it proves that the idea of a
non-trivial critical point at $\theta=\pi$ is not a fluke of the
sphere sigma model, and is true for any $N$. 

The cases with ${\bf Z}_2$ instantons are very similar to the WZW
cases: $\theta$ is not a tunable parameter. In fact, I believe it
is fixed uniquely by the underlying disordered system. Indeed,
the expression for the winding number as an integral over the fields
is precisely of the form of the WZW term. The deep connection between
anomalies and theta terms was discussed in \cite{wieg}.

There are three models with a Pruisken phase, roughly corresponding to
the three kinds of Lie groups $U(N)$, $Sp(2N)$ and $O(2N)$. There are
only two models with ${\bf Z}_2$ instantons, roughly corresponding to
$U(N)$ and $O(2N)$ type, flowing to $U(N)_1$ and $O(2N)_1$ when
$\theta=\pi$. It is logical to ask if there a sigma model with
$Sp(2N)$ symmetry resembling the latter two. From the replica point of
view, it is clearly the sigma model in class $C$, namely
$Sp(2N)/U(N)$. The reason I view this as analogous is that 
the sigma model has action (\ref{pcm}), where
$T$ is unitary and symmetric like the other two, but with the
additional restriction that tr$(JT) =0$, where $J$ is the $2N\times
2N$ matrix
$$J=\pmatrix{0&I\cr -I &0\cr},$$ where $I$ is the $N\times N$ identity
matrix. One can presumably obtain the $Sp(2N)/U(N)$ sigma model from
the $SU(2N)/SO(2N)$ model by pertubing by something like $\lambda \int
(\hbox{tr}\,JT)^2$. This breaks the global $SU(2N)$ symmetry to
$Sp(2N)$. The question is if when $\theta=\pi$, there remains a
non-trivial critical point after perturbing. For $N=1$, the two sigma
models are the same ($Sp(2)=SU(2)$) so obviously the fixed points are
the same. For $N>1$, any non-trivial fixed point in $Sp(2N)/U(N)$
should be a perturbation of the $\theta=\pi$ fixed point of
$SU(2N)/SO(N)$, namely $SU(2N)_1$, which has central charge $2N-2$.
If there is a non-trivial critical point of $Sp(2N)/U(N)$ when
$\theta=\pi$, it must have central charge less than $2N-2$, which
leaves only $Sp(2N)_1$. Obviously, this is not a proof there is such a
point, and moreover, it does not say if this behavior can be continued
to $N\to 0$. Nevertheless, if this correspondence holds, it predicts
that in class $C$ the density of states $\rho(E) \propto E^{1/7}$
\cite{NTW,FK}. This agrees with the exact result of \cite{ilya}
derived from the map onto percolation. It also predicts that there is
another relevant operator of positive dimension $5/4$, which is
thermal operator in percolation and the operator corresponding to
moving off the critical point in the network model. These numbers also
can be found from the analogous supersymmetric approach \cite{andre}.
While I think the agreement of dimensions is not a coincidence, this
hardly proves that the non-trivial fixed point in class $C$ is
$Sp(2N)_1$. It would be much more convincing if a correlator could be
computed and shown to be equivalent correlator in percolation.

So this section has two morals virtually identical to those in the
last:

\begin{list}
\medskip
\item{1.} All the available evidence suggests that when $\theta=\pi$, there
is a non-trivial fixed point, in support of Pruisken's scenario.
\item{2.}  For models with ${\bf Z}_2$ instantons, the underlying disordered
system should determine if $\theta=\pi$ or $0$.
\end{list}
\medskip
\bigskip
I want to add a third moral:

\begin{list}
\medskip
\item{3.} Relevant operators may not always be relevant. 
\end{list}
\medskip
\noindent
What I mean by the last is best illustrated by an example, following
\cite{Affleck}. Consider the principal chiral model on $SU(2)$ with
the action (\ref{pcm}), with the field $T$ taking values in
$SU(2)$. Now add a $k=1$ WZW term, (\ref{wzw}) with $h=T$. As noted
before, this causes a flow to the stable fixed point $SU(2)_1$. Say in
addition to adding the WZW term, I also add a term
$\lambda(\hbox{tr}\, T)^2$. Around the trivial fixed point, this is a
relevant perturbation, breaking the chiral symmetry but not the
diagonal $SU(2)$. One might think it wrecks the flow to the $SU(2)_1$
(chirally-invariant) fixed point. However, it does not necessarily. An
$SU(2)$ matrix $T$ can be rewritten as 
$$\pmatrix{n_0 + i n_1& n_2 + i n_3\cr -n_2 + in_3 & n_0 -in_1\cr}$$
where the otherwise-free parameters must satisfy
$(n_0)^2+(n_1)^2+(n_2)^2+(n_3)^2=1$. The chiral-symmetry-breaking
perturbation is $\lambda (n_0)^2$. For $\lambda$ large, its effect is
to force $n_0=0$, leaving $(n_1)^2+(n_2)^2+(n_3)^2=1$. This is the
sphere: this relevant pertubation turns the principal chiral model
into the sphere sigma model. The WZW term turns into the theta term of
the sigma model, with $\theta=k\pi$. The result discussed above shows
that if $k$ is an odd integer, the presence of both the WZW term and
the chiral-symmetry-breaking perturbation does not result in a massive
theory: one ends up at the $SU(2)_1$ fixed point! For $k=1$, one ends
up exactly where one would have otherwise, although the flow does
reach the $SU(2)_1$ fixed point from a different (chirally
non-invariant) direction. In fact one can see directly at the
$SU(2)_1$ fixed point that all fields
$T_{\alpha\beta}T_{\gamma\delta}$ operator is irrelevant there
\cite{KZ}. However, I think this is a useful moral for the situation
with an unknown low-energy fixed point: just because there is a
relevant operator at the trivial fixed point doesn't necessarily mean
it will always be relevant.

\bigskip
I have discussed how a non-trivial fixed point can
appear in a two-dimensional replica sigma model.  These are summarized
in Table 2. Every universality class save one has at least one kind of
possible non-trivial critical point. Ironically, the only one that
does not is Anderson's original problem of free electrons with
disorder!

\bigskip\bigskip 

I would like to thank Robert Konik for many conversations and for
collaborating on \cite{FK}. I have benefitted enormously from
conversations and correspondence with John Chalker, Ilya Gruzberg,
Andreas Ludwig, Christopher Mudry, Chetan Nayak, T. Senthil, Ben
Simons, Martin Zirnbauer and especially Nick Read. I thank them for
their patience in explaining most of this work to me. I would also
like to thank Alexei Tsvelik for organizing such a nice conference.
This work was supported by a DOE OJI Award, a Sloan Foundation
Fellowship, and by NSF grant DMR-9802813.

\vfill\eject


\begin{thebibliography}{99}
\bibitem{me} P. Fendley, ``Integrable sigma models with $\theta=\pi$'',
to be written up soon. Really.
\end{thebibliography}
\end{document}